\documentclass[onecolumn,preprint]{biophys}
\usepackage{amssymb,amsfonts,amsthm,bm,amsmath}
\usepackage[round,numbers,sort&compress]{natbib}
\usepackage[ansinew]{inputenc}

\jno{xxxx}
\gridframe{N}
\cropmark{Y}
\doi{doi: 10.1529/biophysj.xxxx}

\begin{document}

\title{PDCD5 interacts with p53 and functions as a regulator of p53 dynamics in the DNA damage response}

\author{Changjing Zhuge$^a$, Xiaojuan Sun$^b$,  Yingyu Chen$^c$,   Jinzhi Lei$^d$}

\address{$^a$ College of Sciences, Beijing Forestry University, 35 Tsinghua East Road, Beijing 100083, China; \\
$^b$ School of Science, Beijing University of Posts and Telecommunications, 10 Xitucheng Road, Beijing 100876, China;\\
$^c$  Laboratory of Medical Immunology, School of Basic Medical Science, Peking University, 38 Xueyuan Road, Beijing 100083, China;\\ 
$^d$ MOE Key Laboratory of Bioinformatics, Zhou Pei-Yuan Center for Applied Mathematics, Tsinghua University, Beijing 100084, China}

\begin{abstract}
{The tumor suppressor p53 plays a central role in cell fate decisions after DNA damage. Programmed Cell Death 5 (PDCD5) is known to interact with the p53 pathway to promote cell apoptosis. Recombinant human PDCD5 can significantly sensitize different cancers to chemotherapies. In the present paper, we construct a computational model that includes PDCD5 interactions in the p53 signaling network and study the effects of PDCD5 on p53-mediated cell fate decisions during the DNA damage response. Our results revealed that PDCD5 functions as a co-activator of p53 that regulates p53-dependent cell fate decisions via the mediation of p53 dynamics. The effects of PDCD5 are dose-dependent such that p53 can display either sustained or pulsed dynamics at different PDCD5 levels. Moreover, PDCD5 regulates caspase-3 activation via two mechanisms during the two phases of sustained and pulsed p53 dynamics. This study provides insights regarding how PDCD5 functions as a regulator of the p53 pathway and might be helpful for increasing our understanding of the molecular mechanisms by which PDCD5 can be used to treat cancers. }

{Submitted xxxxxx, and accepted for publication xxxxxx.}
{*Correspondence: jzlei@tsinghua.edu.cn\\
Address reprint requests to Jinzhi Lei, Zhou Pei-Yuan Center for Applied Mathematics, Tsinghua University, Beijing 100084, China\\
Editor:
}
\end{abstract}

\maketitle

\markboth{aaa}{PDCD5 interacts with the p53 pathway}

\def\ptf{[\mathrm{p53}]}
\def\mdmc{[\mathrm{Mdm2_{cyt}}]}
\def\mdmn{[\mathrm{Mdm2_{nuc}}]}
\def\killer{[\mathrm{killer}]}
\def\cytoc{[\mathrm{CytoC}]}
\def\casp{[\mathrm{C3}]}
\def\pdcd{[\mathrm{PDCD5}]}
\def\atm{[\mathrm{ATM}]}
\def\arrester{[\mathrm{arrester}]}
\def\ATM{$\mathrm{ATM}^*$}
\def\p53{$\mathrm{p53}^*$}

\newpage

\section*{INTRODUCTION}

The tumor suppressor p53 plays a central role in cellular responses to various stresses, such as oxidative stress, hypoxia, telomere erosion and DNA damage \cite{Levine2009,Junttila2009}. As a powerful transcription factor, p53 primarily functions by inducing the transcription of many different downstream genes, including p21/WAF1/CIP1 and GADD45, which are involved in cell cycle arrest, and PUMA, Bax and PIG3, which induce apoptosis \cite{Hollstein1991,Laptenko2006,Hanahan2000}. p53 can also control apoptosis through a transcription-independent mechanism \cite{Mihara2003}. Fine control of p53 activity is crucial for proper cellular responses.

In unstressed cells, p53 is maintained at low levels via interactions with E3 ubiquitin ligases, such as Mdm2 \cite{Kubbutat1997}, Pirh2 \cite{Leng2003}, COP1 \cite{Dornan2004} and ARG-BP1 \cite{Chen2005}, which mediate p53 degradation through the ubiquitin-proteasome pathway. Under stressed conditions such as DNA damage, p53 is stabilized and activated to induce the expression of downstream target genes. This process leads to different cellular responses such as cell cycle arrest and apoptosis; the former facilitates DNA repair and promotes cell survival, whereas the latter provides an efficient way to remove damaged cells \cite{Rich2000}. These processes are tightly controlled by the binding partners and post-translational modifications \cite{Meek2009}. For example, upon the occurrence of DNA double-strand breaks (DSBs), the DSB detector ATM is activated and induces the phosphorylation of p53 and Mdm2 \cite{Bakkenist2003, Kitagawa2005,Prives1998,Stommel2004}. Phosphorylation of p53 and Mdm2 inhibits Mdm2-mediated p53 degradation and therefore stabilizes p53. p53 can be phosphorylated or acetylated at multiple sites by different protein kinases, and its stability and sequence-specific DNA binding activity are modulated through these processes \cite{Bode2004}. Phosphorylation at Ser15 by ATM/ATR leads to cell cycle arrest \cite{Abraham2001}, whereas further phosphorylation at Ser46 by HIPK2 promotes the expression of pro-apoptotic genes such as \textit{p53AIP1} \cite{Pomerening2005}. Acetylation of p53 at Lys120 by Tip60 induces the expression of pro-apoptotic genes such as \textit{bax} and \textit{puma} \cite{Tang2006}.

Programmed Cell Death 5 (PDCD5; formerly referred to as TF-1 cell apoptosis-related gene 19 (TFAR19)) is known to promote apoptosis in different cell types in response to various stimuli and also to enhance TAJ/TROY-induced paraptosis-like cell death \cite{Liu1999,Wang2004}. PDCD5 is rapidly upregulated following apoptotic stimuli and translocates from the cytoplasm to the nucleus during early apoptosis \cite{Chen2001}. Decreased expression of PDCD5 has been detected in various human tumors, including lung cancer \cite{Spinola2006}, gastric cancer \cite{Yang2006}, chronic myelogenous leukemia \cite{Ruan2006}, prostate cancer \cite{Du2009}, epithelial ovarian carcinoma \cite{Zhang2011}, astrocytic glioma \cite{Li2008} and chondrosarcoma \cite{Chen2010}. Moreover, the restoration of PDCD5 with recombinant protein or an adenovirus expression vector can significantly sensitize different cancers to chemotherapies \cite{Ruan2008,Chen2010,Shi2010,Wang2009}. Thus, PDCD5 likely plays a critical role in multiple tissues during tumorigenesis. However, the molecular mechanisms that underlie the function of PDCD5 during cell growth, proliferation and apoptosis remain largely unclear.

Previous experiments have demonstrated that PDCD5 is apparently upregulated in cells following apoptotic stimulation \cite{Liu1999}, enhances caspase-3 activity by modulating Bax translation from the cytosol to the mitochondrial membrane \cite{Chen2006a}, interacts with Tip60 to enhance histone acetylation and p53 acetylation at Lys120, and promotes the expression of Bax \cite{Xu2009}. Recently, novel evidence indicated that PDCD5 is a p53 regulator during gene expression and the cell cycle \cite{Xu2012a}. It was shown that PDCD5 interacts with the p53 pathway by inhibiting the Mdm2-mediated ubiquitination and nuclear exportation of p53 and that knockdown of PDCD5 can decrease the ubiquitination level of Mdm2 and attenuate the expression and transcription of p21. Hence, upon DNA damage, PDCD5 can function as a co-activator of p53 to regulate cell cycle arrest and apoptosis.

Many computational models have been constructed to investigate the mechanism of the p53-mediated cell-fate decision \cite{Bar-Or2000, Mihalas2000,Tiana2002,Michael2003,Ma2005,Geva-Zatorsky2006,Zhang2009, Zhang2010a, ZhangPNAS2011, Zhang2012, Tian2012, Batchelor2008, Kim2013}. In these models, the p53/Mdm2 oscillation is highlighted as important to the cell-fate decision following DNA damage. Integrated models of the p53 signaling network have been established to study the process of cell fate decision in response to DNA damage \cite{Zhang2009,Zhang2010a,ZhangPNAS2011,Zhang2012}. These models advance the understanding of the dynamics and functions of the p53 pathway in the DNA damage response. In \cite{Zhang2009}, it has been suggested that the decision between the cell fates of survival and death might be determined by counting the number of p53 pulses. In \cite{ZhangPNAS2011}, the two feedback loops of ATM-p53-Wip1 and p53-PTEN-Akt-Mdm2 are combined in the p53 signaling network. A two-phase p53 response has been shown in this model; pulses occur during DNA repair and are sustained at a high level that triggers apoptosis if the damage cannot be fixed after a crucial number of p53 pulses. Furthermore, dynamical analysis has shown that the ATM-p53-Wip1 loop is essential for the generation of the p53 pulses and that the PTEN level determines whether p53 acts as a pulse generator or a switch. Despite extensive studies of the p53 pathway, little work has been focused on modeling the PDCD5 interactions. The first model of PDCD5-regulated DNA damage decisions was established by \cite{Zhuge2011}. Two known pathways were considered in this model: the interaction of PDCD5 with Tip60 in the nucleus, and the regulation of Bax translocation in the cytoplasm. This model revealed that the cytoplasmic pathway plays an important role in PDCD5-regulated cell apoptosis \cite{Zhuge2011}. However, how PDCD5 interactions with the p53 pathway affect cell fate decision has not been considered in previous models.

Motived by the above considerations, we constructed a mathematical model of the p53 signaling network with PDCD5 interactions in the present study to examine the effects of PDCD5 on p53-mediated cell fate decisions in response to DNA damage.  The main results of this study suggest that PDCD5 can function as a co-activator of p53 to regulate p53-dependent cell fate decisions by mediating the dynamics of p53. The effects of PDCD5 are dose dependent, and various cell fates can occur for cells with different PDCD5 levels.

\section*{MATERIALS AND METHODS}
\subsection*{Model description}
\begin{figure}[htbp]
 \centering
 \includegraphics[width=8cm]{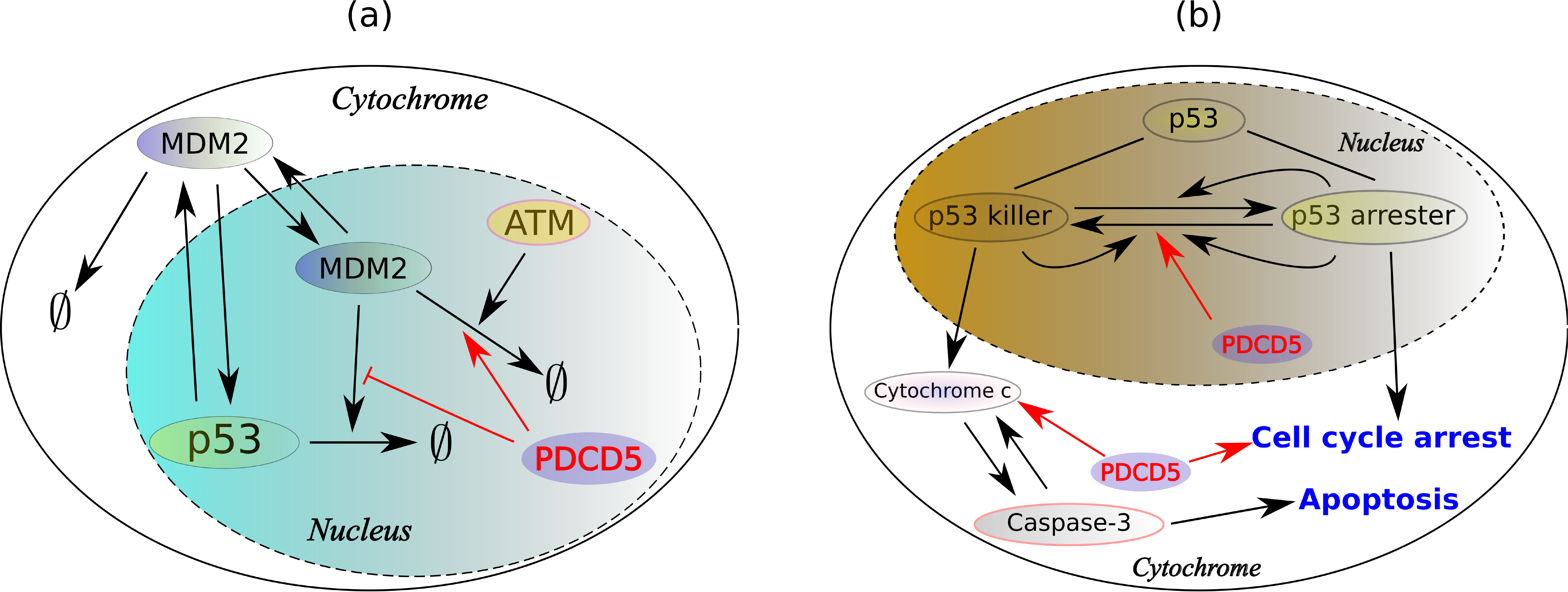}
 \caption{Model of the PDCD5 pathway that regulates p53/MDM2 oscillation and cell fate decision. (a) Model of PDCD5-regulated p53/MDM2 oscillation. (b) Model of the cell fate decision. Red color lines show the PDCD5 interactions.}
 \label{fig:SimplifiedModel}
\end{figure}

Our model was based on p53 responses to DNA damage caused by ionizing radiation (IR) \cite{Zhang2009,ZhangPNAS2011} and PDCD5 interactions with the p53 pathway \cite{Xu2012a} (Figure \ref{fig:SimplifiedModel}). In the model, the cell fate decision following DNA damage is mediated by p53 pulses through the p53-Mdm2 oscillator, and PDCD5 interacts with p53 and functions as a positive regulator in the p53 pathway.

An integrated model with four modules for the p53 signaling network has been developed by \cite{Zhang2009,ZhangPNAS2011}. This model includes the following processes: DNA repair, ATM switch, p53-Mdm2 oscillation, and cell fate decision. When a cell is exposed to IR, a certain number of DSBs are generated in the cell and induce the formation of DSB repair-protein complexes (DSBCs), and the repair process ensues. Subsequently, DSBCs promote the conversion of inactive ATM monomers to active forms \cite{Bakkenist2003} such that active ATM (\ATM) becomes dominant after exposure to IR.

After the activation of ATM, the p53 level exhibits a series of pulses due to the feedback loops in the p53-Mdm2 oscillator. The protein p53 and its negative regulator Mdm2 are the core proteins in this oscillator (Figure \ref{fig:SimplifiedModel}a). In the nucleus, p53 is activated by \ATM in two ways. First, \ATM promotes the phosphorylation of p53 on Ser-15 \cite{Prives1998} and accelerates the degradation of Mdm2 through phosphorylation \cite{Stommel2004}. Thus, \ATM induces a conversion of p53 from the inactive state to the active state (\p53) \cite{Stommel2004}. Second, \p53 is deactivated at a basal rate. Here, only \p53 can induce the production of $\mathrm{Mdm2}_{\mathrm{cyt}}$, which in turn promotes the translation of \textit{p53} mRNA in the cytoplasm \cite{Yin2002}. In undamaged cells, p53 levels are kept low by Mdm2 through the negative-feedback between \p53 and $\mathrm{Mdm2}_{\mathrm{nut}}$. After damage, the p53-Mdm2 complex is dissociated due to activation of p53 by \ATM, and the levels of \p53 and $\mathrm{Mdm2}_{\mathrm{cyt}}$ increase abruptly through the positive feedback between \p53 and $\mathrm{Mdm2}_{\mathrm{cyt}}$.

In the cell fate decision module, p53 coordinates cell cycle arrest and apoptosis to govern cell fate through the phosphorylation of p53 at distinct sites (Figure \ref{fig:SimplifiedModel}b). The primary phosphorylation of p53 on Ser-15 leads to cell cycle arrest, whereas the further phosphorylation of p53 on Ser-46 promotes expression of pro-apoptotic genes such as \textit{p53AIP1} \cite{Oda2000}. The above two forms of phosphorylated p53 are termed p53 arrester and p53 killer, respectively \cite{Zhang2009}. There are three feedback loops involved in the conversion between p53 arrester and p53 killer. The p53 arrester-inducible gene \textit{Wip1} can promote the reversion of p53 killer to p53 arrester \cite{Fiscella1997}, while the gene \textit{p53DINP1}, which is induced by both p53 arrester and p53 killer, contributes to the formation of p53 killer \cite{Okamura2001}. p53 arrester induces cell cycle arrest through the transcriptional activation of \textit{p21}, and p53 killer promotes cell death via pro-apoptotic genes such as \textit{p53AIP1}. The over-expression of \textit{p53AIP1} induces the release of cytochrome \textit{c} from mitochondria, and apoptosis rapidly ensues after the activation of caspase-3. A positive-feedback loop between cytochrome \textit{c} and caspase-3 \cite{Kirsch1999,Bagci2006} underlies the apoptotic switch in the model developed by \cite{Zhang2009}.

The PDCD5 protein is weakly expressed in unstressed conditions, and is upregulated in cells upon apoptotic simulation \cite{Liu1999}. Following the onset of an apoptotic stimulus, the cytoplasmic PDCD5 protein level first increases rapidly and forms an inward gradient from the cytoplasm to the nucleus \cite{Liu1999,Chen2001,Xu2009}. This pattern is maintained for a few hours until the initiation of apoptosis, which is paralleled by a rapid translocation of PDCD5 protein from the cytoplasm to the nucleus \cite{Chen2001}. Hence, after DNA damage, nuclear PDCD5 is maintained at an intermediate level  during the DNA repair process.

PDCD5 has been found to up-regulate p53 activity through at least two interactions \cite{Xu2012a}. When U2OS cells were transfected with either control or PDCD5-specific siRNA, p53 protein levels decreased following the knockdown of PDCD5. Simultaneously, the knockdown of PDCD5 failed to influence the levels of p53 mRNA levels, which suggests that PDCD5 enhances the stability of p53 and does not regulate p53 at the transcriptional level \cite{Xu2012a}. Co-localization analysis in U2OS cells further revealed that PDCD5 co-localizes with p53 in the nucleus. Furthermore, NMR experiments indicated that PDCD5 can bind with the N-terminal domain of p53 (p53$_{15-61}$) \cite{Yao2012}, which overlaps with the binding site between p53 (p53$_{15-29}$) and Mdm2 \cite{Schon2004}. When p53 was incubated with whole lysates of HeLa cells that overexpressed Mdm2, p53 strongly bound to Mdm2, but this interaction between p53 and Mdm2 decreased significantly in the presence of recombinant human PDCD5. Moreover, PDCD5 can be pulled down with p53. These results suggest that PDCD5 might disrupt the p53-Mdm2 interaction via the direct interaction between p53 and PDCD5 \cite{Xu2012a}, which is consistent with the results of the NMR study \cite{Yao2012}. Interestingly, PDCD5 has been found to be capable of dose-dependently decreasing the protein level of Mdm2. Knockdown of endogenous PDCD5 could increase the accumulation of Mdm2 and decrease the ubiquitination level of Mdm2 \cite{Xu2012a}. Hence, PDCD5 dissociates the p53-Mdm2 complex promotes Mdm2 degradation. These interactions are shown with red lines in Figure \ref{fig:SimplifiedModel}a.

At the cell fate decision module, PDCD5 in the nucleus interacts with Tip60 to promote the Tip60-induced Lys120 acetylation of p53 (the killer form of p53) \cite{Xu2009}. In the cytoplasm, PDCD5 promotes the translocation of Bax from the cytosol to the mitochondrial outer membrane to induce the release of cytochrome \textit{c} \cite{Chen2006a}. ChIP assays in U2OS cells have shown that PDCD5 might associate with the \textit{p21} promoter to promote transcription activation after DNA damage \cite{Xu2012a}. Knockdown of PDCD5 attenuates the expression and transcription of p21 \cite{Xu2012a}. Hence, PDCD5 in the nucleus increases the transition from p53 arrester to p53 killer and promotes the transcription of p21, and PDCD5 in the cytoplasm up-regulates Bax translocation to increase cytochrome \textit{c} release (Figure \ref{fig:SimplifiedModel}b).

Based on the model simulation reported by \cite{Zhang2009}, active ATM is dominant following IR and the level of \ATM remains mostly constant during the DNA repair process. Hence, in our model, the four modules of the model of \cite{Zhang2009} were simplified to include only the two modules of the p53-Mdm2 oscillator and the cell fate decision, and the levels of \ATM and nuclear PDCD5 were represented by time dependent functions to mimic the DNA repair process. This simplification is acceptable in the current study because we intended to investigate the effects of PDCD5 on p53 dynamics after DNA damage. For a more complete understanding of the effect of PDCD5 on cell fate decisions following DNA damage, an integrated model that incorporates PDCD5 dynamics \cite{Zhuge2011} and the responses of the p53 pathway \cite{Zhang2009,Zhang2011} is certainly required and will be the subject of further studies.

\subsection*{Formulations}

In the formulations, we first simplified the models presented in \cite{Zhang2009,Zhang2011} to a six differentiation equations for the modules of the p53-Mdm2 oscillator and cell fate decision, and a time-dependent \ATM level was introduced for the DNA repair process.

In the p53-Mdm2 oscillator, inactive p53 in the nucleus is degraded rapidly by Mdm2 and was thus assumed to be at quasi-equilibrium in our model. Hence, three components were included in the p53-Mdm2 oscillator: active p53 in the nucleus $\ptf$, Mdm2 in the nucleus $\mdmn$, and Mdm2 in the cytoplasm $\mdmc$. Active p53 promotes the production of $\mathrm{Mdm2}_{\mathrm{cyt}}$, which promotes the translation of \textit{p53} mRNA to produce p53 to form a positive-feedback loop. In the nucleus, active p53 is degraded slowly by weakly binding to $\mathrm{Mdm2_{nuc}}$, and the interaction is disrupted by PDCD5. Mdm2 in the nucleus and cytoplasm can be shuttled between the two compartments at different rates. The degradation of $\mathrm{Mdm2}_{\mathrm{nuc}}$ is promoted by both \ATM and PDCD5. These interactions resulted in differential equations (1)-(3), which are given in the Supporting Material.

In the cell fate decision module, there are two forms of active p53: p53 arrester and p53 killer. p53 arrester and p53 killer transform into each other at different rates that are regulated by their inducible genes \textit{Wip1} and \textit{p53DINP1}. PDCD5 in the nucleus can increase the transition from p53 arrester to p53 killer. p53 killer induces apoptosis through the killer-inducible gene \textit{p53AIP1}, which up-regulates the expression of the pro-apoptotic gene \textit{Bax}. PDCD5 in the cytoplasm enhances Bax translocation and promotes the release of cytochrome \textit{c} from the mitochondria and hence the activation of caspase-3. In our model, we omitted components for the proteins Wip1, p53DINP1, and p53AIP1 and considered the dynamics of p53 killer $\killer$ (the p53 arrester concentration is given by $\arrester = \ptf - \killer$), cytochrome \textit{c} $\cytoc$ and active caspase-3 $\casp$. This process resulted in equations (4)-(6) in the Supporting Material. Here we note that the downstream signal p21 transcription was omitted.

The intensities of \ATM and PDCD5 were included in the model through their incorporation in the equation coefficients (refer to the Supporting Material, Section 1.6). This study intended to study p53 dynamics during DNA repair while both \ATM and PDCD5 remained at high levels \cite{Zhang2009,Zhuge2011}. Hence, the \ATM and PDCD5 levels were described by the predefined functions $A(t)$ and $P(t)$ given by equation (20)-(21) in the Supporting Material. The parameter $P_0$ was introduced for the different PDCD5 levels with $P_0 = 0.8$ for the wild-type cells and $P_0=0.2$ for the siPDCD5 cells. Despite this specificity, extensive simulations showed that the results of the current study were insensitive to different mathematical formulations of these two functions.

Now, the original integrated model of cell fate decision mediated by p53 pulses has been simplified to the above model with six differential equations. With this simplification, all parameters were adjusted to reproduce the p53 pluses utilized by \cite{Zhang2009,Zhang2011}, and the PDCD5 interaction parameters were estimated based on experimental results regarding the degradation of p53 with or without PDCD5 \cite{Xu2012a}. Details of the parameter values are given in Table 1 in the Supporting Material.

\subsection*{Numerical methods}
In the numerical simulations, the model equations were solved numerically using \verb|NDSolve| on the Mathematica 8.0 platform \cite{mathematica8}.

\section*{RESULTS}
\subsection*{PDCD5 regulates p53 dynamics in a dose-dependent manner}

\begin{figure}[htbp]
\centering
\includegraphics[width=8cm]{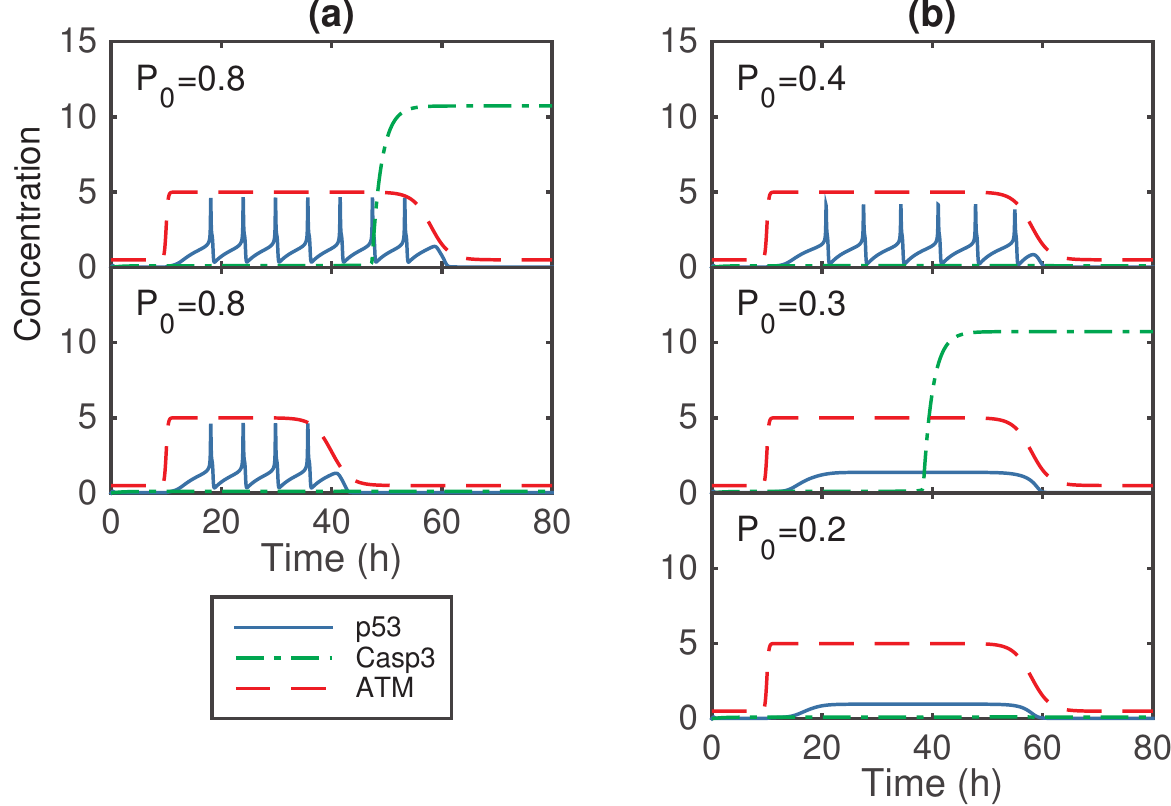}
\caption{Typical p53 dynamics after DNA damage. The DNA repair processes are shown by high ATM levels (red dashed lines), the PDCD5 level was adjusted by $P_{0}$ (shown at each figure panel), and the cell fate was indicated by caspase-3 (green dash dotted lines). (a) Wild-type cells with $P_{0} = 0.8$ and $48 h$ and $30 h$ DNA repair processes, respectively. (b) Cells with various PDCD5 levels and $48 h$ DNA repair processes.}
\label{fig:SamplePaths}
\end{figure}

Signaling dynamics are known to encode and decode cellular informations that controls cellular responses \cite{Purvis:2013}. p53 dynamics can control cell fate decision in response to DSBs, and cells that experience p53 pulses recover from DNA damage, whereas cells that are exposed to sustained p53 signaling frequently undergo senescence \cite{Purvis:2012}. To examine how PDCD5 regulates the dynamics and functions of p53, we performed simulations with various PDCD5 levels. The integrated model showed that cell fate was governed by the number of p53 pulses during DNA repair \cite{Zhang2009}. In our simulations, there were seven p53 pulses when the DNA repair process required $t_{c}=48 h$ and apoptosis was induced by obvious increases in caspase-3 levels. When the DNA process is shortened to $t_{c}=30 h$, there are four p53 pulses and the cells recover to normal growth, while caspase-3 is maintained at a low level (Figure \ref{fig:SamplePaths}a). These results reproduced the p53 dynamics and cell fate decision obtained from the integrated model of \cite{Zhang2009}; therefore, our model simplification is capable of investigating the effects of PDCD5 on p53 dynamics.

To examine the effects of PDCD5 on p53 dynamics, we fixed the DNA repair process at $t_{c}=48 h$ and varied the PDCD5 level to examine the cell response. Increasing $P_{0}$ elicited no changes in cell fate (data not shown), and reducing the PDCD5 level might lead to various p53 dynamics and cell fate (Figure \ref{fig:SamplePaths}b). When $P_{0}=0.4$, there are six p53 pulses during DNA repair and caspase-3 is maintained at low levels so that the cells recover to normal growth. The p53 pulses are repressed if $P_{0}$ is further reduced. When $P_{0}=0.3$, the cells exhibit sustained p53 signaling and caspase-3 increases to levels as high as those observed when $P_{0}=0.8$. When $P_{0}$ is reduced to $0.2$, the p53 level is attenuated and fails to induce cell apoptosis. These results suggest that p53 dynamics and cell fate can be modulated by PDCD5 in a dose-dependent manner. Here we note that when $P_{0}=0.3$, the caspase-3 reaches levels as high as those in the wild-type cells ($P_{0} = 0.8$). However, at this point we cannot conclude that the cells undergo apoptosis \cite{Abraham2004} because some other response pathways not included in the current study, such as cell senescence, can be triggered by sustained p53. In \cite{Purvis:2012}, it was suggested that a proper stimulus with Nutlin-3 can induce sustained p53. Our simulations predict that a proper dose of PDCD5 can also result in sustained p53 after DNA damage. This prediction requires experimental confirmation.

\begin{figure}[htbp]
 \centering
 \includegraphics[width=6cm]{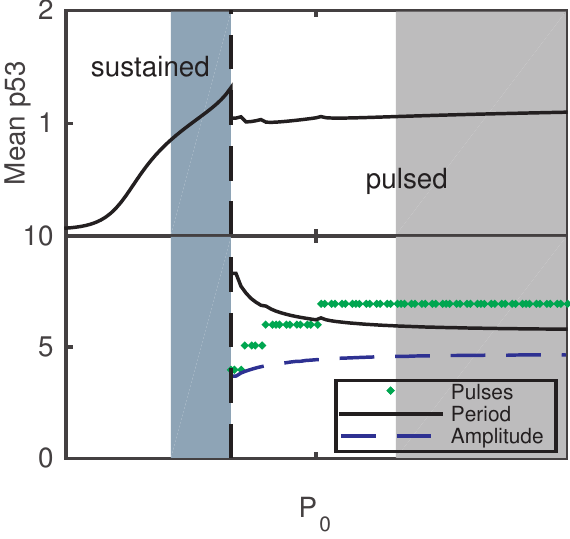}
 \caption{The effects of PDCD5 on p53 dynamics. Here we set $t_{c}=48 h$ and varied $P_{0}$ from $0$ to $1$. Upper panel, the mean p53 level over the DNA repair process. Bottom panel, the p53 pulse numbers, period, and amplitude during DNA repair. The shadowed regions indicate the $P_{0}$ range at which caspase-3 is activated. The vertical dashed line separates $P_{0}$ into sustained or pulsed p53 dynamics. }
 \label{fig:dyns}
\end{figure}

To further investigate the effects of PDCD5 on p53 dynamics, we altered $P_{0}$ over the wider range of $0$ to $1$. There is a threshold of $P_{0}=0.33$ that corresponds to a Hopf bifurcation of the p53 oscillator module (dashed line in Figure \ref{fig:dyns}) such that p53 is sustained when $P_{0}$ is less than the threshold and p53 is pulsed when $P_{0}$ is above the threshold. In both regions of either sustained or pulsed p53, caspase-3 is activated when $P_{0}$ is relatively large (shadows in Figure \ref{fig:dyns}) but with different  mechanisms. In the sustained region, during DNA repair, the p53 level increases with PDCD5 to induce caspase-3 activation through saddle node bifurcation; however, in the pulsed region, p53 oscillates while the mean value is nearly unchanged with the PDCD5 level, and hence, caspase-3 is activated by an alternative mechanism (see Figure \ref{fig:dyns} upper panel, to be detailed below). In the pulsed region, the period decreases with $P_{0}$ such that the pulse number increases from $4$ to $7$ when $P_{0}$ varies from $0.33$ to $1$, and the amplitude slightly increases with $P_{0}$ (Figure \ref{fig:dyns}, bottom panel). We note that caspase-3 is either active or not when there are $7$ pulses, which indicates that the p53 pulse number alone is insufficient to determine cell fate. These results suggest that PDCD5 regulates the p53 dynamics with different mechanisms in the p53 sustained and pulsed regions that are separated by a Hopf bifurcation of the p53 oscillator module.

\subsection*{PDCD5 regulates caspase-3 activation by two mechanisms}

To investigate the mechanism by which PDCD5 regulates caspase-3 activation, we considered the cell fate decision module, which can be described by the dynamics of cytochrome \textit{c} and caspase-3 given by equations (22)-(23) in the Supporting Material. The caspase-3 dynamics are controlled by the cytochrome \textit{c} release rate $v_0$, which is dependent on PDCD5 and p53 killer. We performed bifurcation analysis for the equations to seek the mechanisms by which PDCD5 induces caspase-3 activation.

\begin{figure}[htbp]
\centering
\includegraphics[width=8cm]{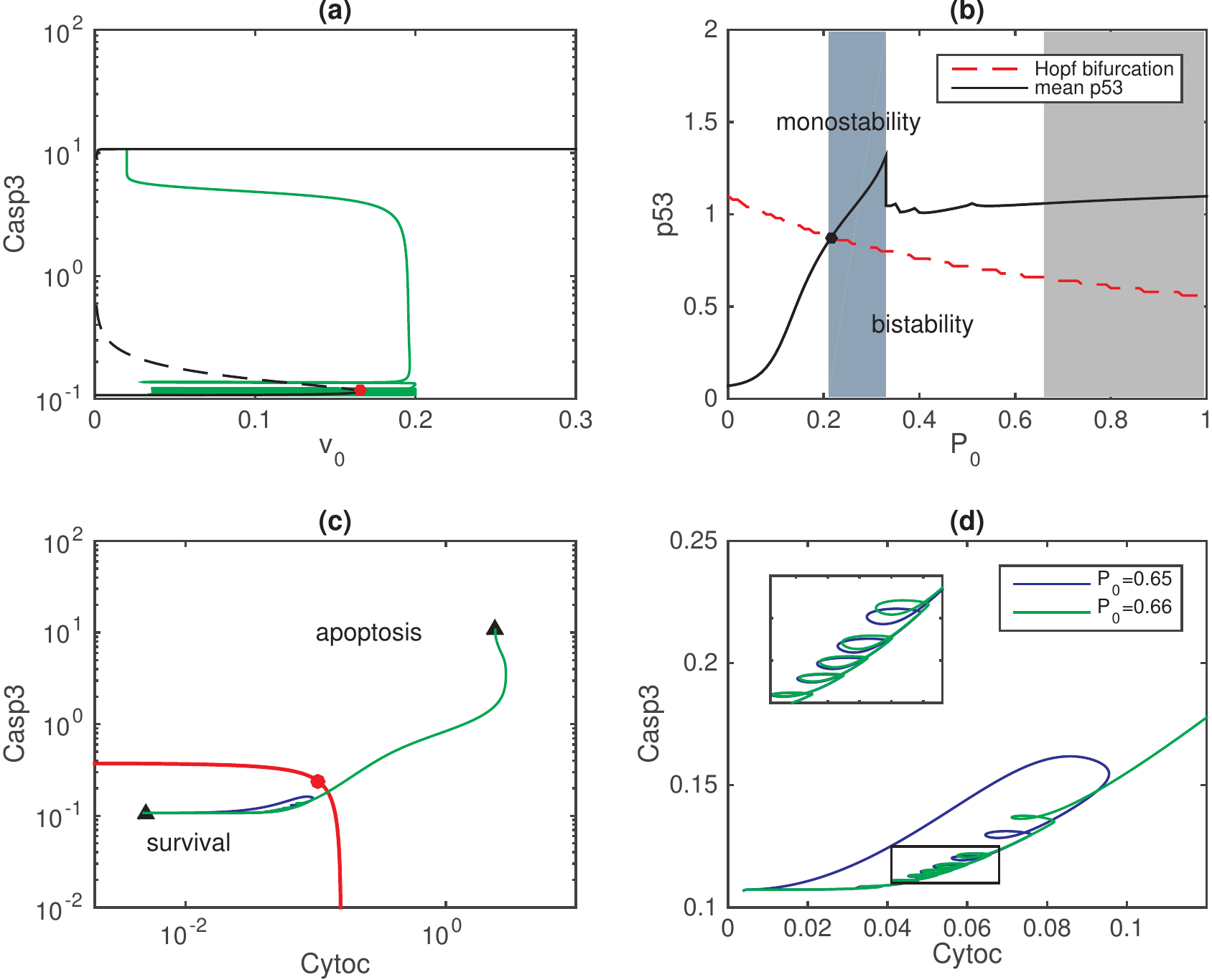}
\caption{Bifurcation analysis of the cell fate decision module. (a) Black shows the dependence of the steady state caspase-3 level on the cytochrome \textit{c} release rate $v_0$ (unstable steady states are marked by the dashed line). Green shows the solution of the original model with $P_0 = 0.66$ and $t_c=48 h$ (same as in (c) and (d)). (b) The superposition of the dependence of the mean p53 on $P_0$ and the curve corresponds to saddle note bifurcation. (c) Phase plane analysis with $v_0 = 0.02$. The black triangles indicate stable steady states, and the red dot indicates the unstable steady state (saddle point). The red curve shows the stable manifolds of the saddle point that divide the phase plane into two regions. The green curve shows the solution of the original model with $P_0 = 0.66$ and $t_c=48 h$ (the cell proceeds to apoptosis), and the blue indicates the solution with $P_0=0.65$ (cell survival). (d) Enlargement of the two solutions shown in (c). The inset enlarges the square region.}
\label{fig:bif}
\end{figure}

In the case of sustained p53, the rate $v_{0}$ is a constant such that the cell fate module exhibits either bistability for small $v_0$ or monostability for larger $v_0$ (Figure  \ref{fig:bif}a). The saddle node bifurcation defines a critical release rate (red point at Figure \ref{fig:bif}a) such that when $v_0$ increases above the critical rate, the low-caspase-3 state vanishes and the system switches to the state of caspase-3 activation. This critical rate defines a critical curve of p53 versus $P_{0}$ through equation (21) in the Supporting Material as shown by the red dashed line in Figure \ref{fig:bif}b. To test whether PDCD5 induces caspase-3 activation during sustained p53 through saddle note bifurcation, we superpose this critical curve with the dependence of the mean p53 on $P_0$ (Figure \ref{fig:dyns}). The two curves meet at a point at which $P_0$ is at the critical value for inducing caspase-3 activation (Figure \ref{fig:bif}b). This consistence indicates that PDCD5 induces caspase-3 through saddle node bifurcation in the region of sustained p53.

When p53 is pulsed, equations (22)-(23) in the Supporting Material are time-dependent and have periodic coefficients. In this case, caspase-3 is not always active even when the mean p53 level is above the critical value (Figure \ref{fig:bif}b). Hence, some other mechanism must be at work to induce caspase-3 activation with increases in the PDCD5 level $P_{0}$. In the absence of DNA damaging stimuli, the caspase-3 dynamic system has two stable steady states of survival and apoptosis (Figure \ref{fig:bif}c). The two stable states are separated by stable manifolds of an unstable steady state (red curve in Figure \ref{fig:bif}c). After DNA damage, starting from the survival state, a cell transitions to the apoptotic state with an increase in the amount of cytochrome \textit{c} released and a rapid caspase-3 activation (Figure \ref{fig:bif}c). The simulations showed that the final cell fate is determined by whether the solution trajectory shown by Figure \ref{fig:bif}c crosses the boundary between the survival and apoptosis regions.  During DNA repair, p53 is oscillating and switching between pro-apoptosis and pro-survival (Figure \ref{fig:bif}a) such that the released cytochrome \textit{c} accumulates at each p53 pulse. Simulations showed that PDCD5 can increase the accumulation of cytochrome \textit{c} and hence induce caspase-3 activation by promoting the solution trajectory to cross the boundary curve during DNA repair (Figure \ref{fig:bif}d). These results indicate that PDCD5 promotes cell apoptosis (caspase-3 activation) as a co-activator of p53 that accelerates cytochrome \textit{c} release during the region of pulsed p53. This observation highlights the crucial role of PDCD5 in the cytoplasm and is in agreement with our previous study \cite{Zhuge2011}.

\begin{figure}[htbp]
\centering
\includegraphics[width=6cm]{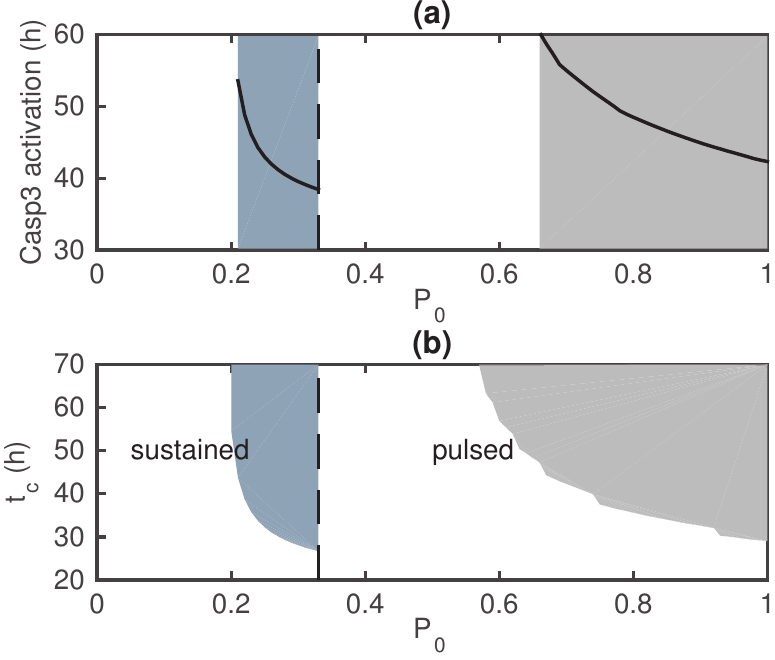}
\caption{Dependence of cell fate on the DNA repair time $t_c$ and the PDCD5 level $P_0$. (a) The dependence of caspase-3 activation time on $P_{0}$, here $t_c = 48 h$. (b) The shaded area shows the DNA repair duration for each $P_{0}$ required to activate caspase-3 with sustained and pulsed p53. }
\label{fig:P0}
\end{figure}

The above analyses indicated that PDCD5 promotes caspase-3 activation by accelerating cytochrome \textit{c} release. Consequently, the time of  caspase-3 activation should decrease with increases in PDCD5 levels. This notion was confirmed by our simulations as shown by Figure \ref{fig:P0}a. Next, we asked whether the PDCD5 level required to induce apoptosis is related to the duration of DNA repair. We changed both the DNA repair duration $t_c$ and $P_0$ to examine cell responses. The simulations showed that caspase-3 activation is induced only when the DNA repair duration is sufficiently large ($t_{c} > 29 h$ in this study). When $t_{c}$ is sufficiently large, the critical PDCD5 level $P_{0}$ required ot induce caspase-3 activation decreases with increases in $t_{c}$ in both regions of sustained and pulsed p53 (Figure \ref{fig:P0}b). We note that the boundary value of $P_{0}$ at which p53 transits from sustained to pulsed dynamics is independent of $t_{c}$ (Figure \ref{fig:P0}b dashed line). This $P_{0}$ value is determined by the Hopf bifurcation of the p53/Mdm2 oscillation module.

\subsection*{The cytoplasm pathway and the regulation of Mdm2-mediated p53 degradation are the primary effects of PDCD5 on the cell fate decision}

\begin{figure}[htbp]
\centering
\includegraphics[width=6cm]{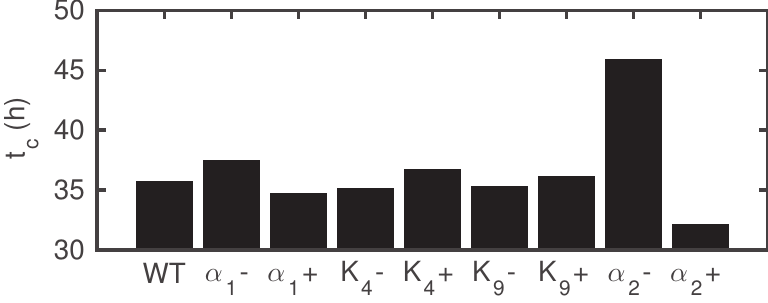}
\caption{Dependence of the critical duration of DNA repair on the model parameters. The WT cells took the value listed in the Supporting Material; for the other cells, each of the labeled parameters was decreased ($-$) or increased ($+$) by $20\%$.}
\label{fig:par}
\end{figure}

PDCD5 interacts with the p53 pathway in multiple ways; it stabilizes p53 by disrupting the p53-MDM2 interaction, enhances Mdm2 degradation, and promotes the Tip60-induced Lys120 acetylation of p53. In the cytoplasm, PDCD5 enhances Bax translocation and promotes the release of cytochrome \textit{c}. PDCD5 regulates the cell fate decision via the combination of these multi-site interactions. To investigate which role is the most essential for the regulatory function of PDCD5, we altered the strengths of each interaction and examined the DNA repair time required to induce cell apoptosis. Four parameters were considered: $\alpha_{1}$ for PDCD5 disrupting the p53-MDM2 interaction, $K_{4}$ for enhancing the Mdm2 regulation, $K_{9}$ for promoting the p53 killer transformation, and $\alpha_{2}$ for the cytoplasm pathway (Figure \ref{fig:par}). The results revealed that cell fate is sensitive to changes in $\alpha_{2}$, i.e., the PDCD5 interactions with the cytoplasm pathway. Among the functions of PDCD5 in the p53 pathway, changes in the strength with which PDCD5 disrupts the p53-Mdm2 interaction ($\alpha_{1}$) result the greatest changes in the DNA repair duration required to induce apoptosis. These results show that the cytoplasm pathway is essential for the regulatory function of PDCD5, which agrees with previous studies \cite{Zhuge2011}, and the PDCD5-regulated disruption of the Mdm2-mediated p53 degradation is important among the interactions of PDCD5 with the p53 pathway. These observations provide poptential options for killing cancer cells in clinical treatments, which are discussed below.

\subsection*{Effects of PDCD5 disrupting Mdm2-mediated p53 degradation}

\begin{figure}[htbp]
\centering
\includegraphics[width=8cm]{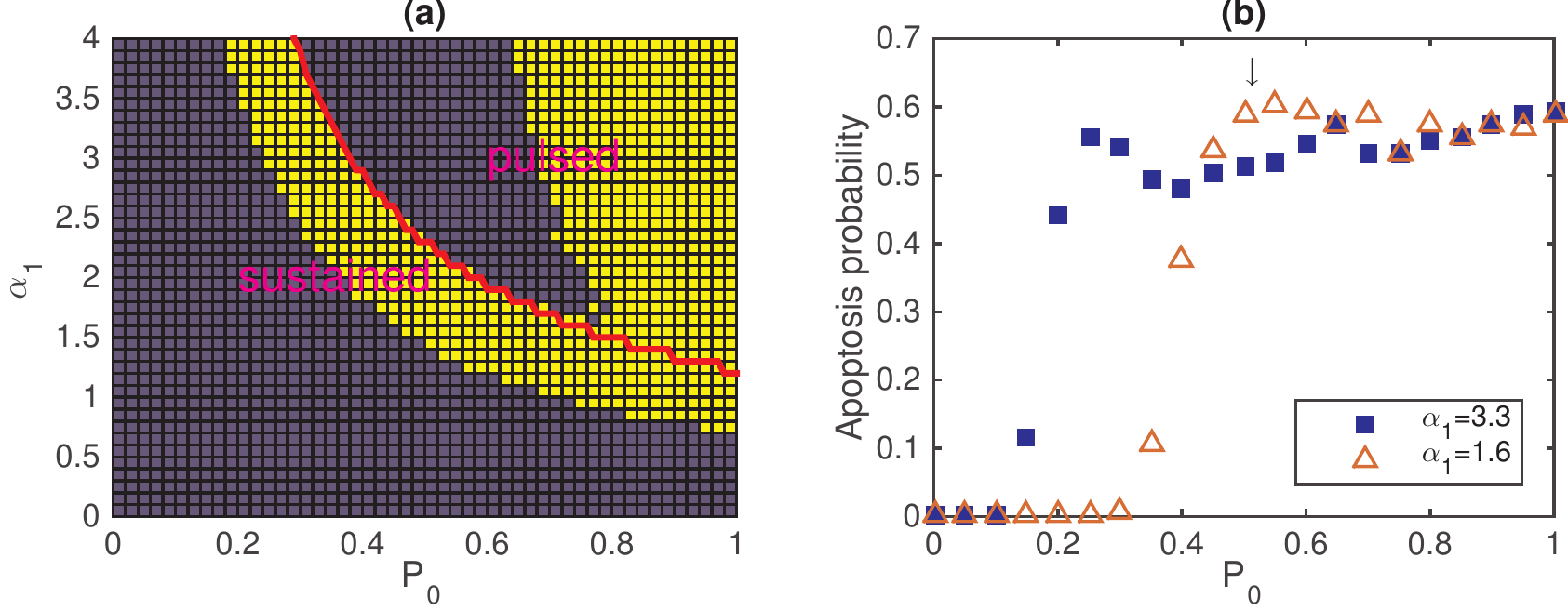}
\caption{Dependence of the cell fate decision on $\alpha_1$. (a) The cell fate decision with various parameters ($\alpha_1$ and $P_0$). Yellow shows the region with caspase-3 activity, and the regions with sustained and pulsed p53 are separated by the red curve. (b) Apoptosis probabilities obtained from mutlicell simulations with varying $P_0$ and different $\alpha_1$ values. Here, $t_c = 48 h$ in all simulations.}
\label{fig:a1}
\end{figure}

PDCD5 disrupts Mdm2-mediated p53 degradation via a direct interaction with p53 \cite{Yao2012,Xu2012a}. Biologically,  the parameter value $\alpha_1$ is adjustable if we can modify the binding affinity between PDCD5 and p53 via methods such as single molecule engineering. To further explore how various affinities ($\alpha_1$) affect the cell fate decision, we varied $\alpha_1$ and $P_0$ to examine the p53 dynamics and caspase-3 activity (Figure \ref{fig:a1}a). The results revealed that caspase-3 was activated only when $\alpha_1 P_0 > 0.7$, and that the two parameter regions with either sustained p53 or pulsed p53 were separated by the curve $\alpha_1 P_0 = 1.1$. For small $\alpha_1$ values ($\alpha_{1} < 1.7$), pulsed p53 always yielded caspase-3 activity; however, when $\alpha_1$ was large,  pulsed p53 did not necessarily imply caspase-3 activity when $P_0$ was not sufficiently large (see Figure \ref{fig:SamplePaths}b).

To further investigate the cellular apoptosis probability, we applied the method of multicell simulation\cite{Zhuge2011} to simulate a group of $10^4$ cells  each of which had parameters that were randomly chosen from a range of $\pm 20\%$ away from their default values. The cell fate of apoptosis was marked by caspase-3 activation, and the fraction of apoptotic cells provided the apoptosis probability. Figure \ref{fig:a1}b shows the apoptosis probabilities with $P_{0}$ from $0$ to $1$ and strong and weak affinity $\alpha_{1}$ values. The results suggested that when PDCD5 is low, the significant increase in the expression of $\alpha_{1}$ can significantly enhance cell apoptosis. We note that when $P_{0}$ takes an intermediate value ($P_{0}\approx 0.5$, arrow at Figure \ref{fig:a1}b), increasing $\alpha_{1}$ tends to decrease the probability of apoptosis. This counterintuitive result was due to the possibility that a cell can display pulsed p53 without activation of caspase-3. These results provide possible directions for interfering with the binding affinity of PDCD5 and p53 to modulate cell fate decisions in clinical treatments.

\section*{DISCUSSION}
PDCD5 is known to interact with p53 and functions as a regulator in the p53 pathway during responses to DNA damage. In the present study, we constructed a mathematical model of the p53 signaling network with the interactions of PDCD5. The model was based on the integrated model of the p53 signaling network that was previously proposed \cite{Zhang2009,Zhang2012}, and the interactions of PDCD5 with the p53/Mdm2 oscillator and cell fate decision were included in accordance with recent observations \cite{Xu2012a}. The computational model consisted of two p53/Mdm2 oscillator and cell fate decision modules. The DNA repair process was represented by increases in active ATM and PDCD5 concentrations, both of which were given by pre-defined time-dependent functions and were incorporated into the equation coefficients.

The model simulations showed that PDCD5 can modulate the cell fate decision by mediating p53 dynamics in a dose-dependent manner such that p53 can display either sustained or pulsed dynamics in cells with different levels of PDCD5. Moreover, PDCD5 regulates caspase-3 activation via two mechanisms that operate in the two regions of sustained and pulsed p53 dynamics. We found that the cell fate decision is sensitive to the cytoplasm pathway of PDCD5, which agrees with the results of our previous studies \cite{Zhuge2011}. Moreover, the PDCD5-regulated disruption of Mdm2-mediated p53 degradation is also important for the interaction of PDCD5 with the p53 pathway, which suggests that it is possible to modulate cell fate decision via interference in the binding affinity between PDCD5 and p53. 

This study sought to investigate the effects of PDCD5 on p53 dynamics following DNA damage. A more comprehensive analysis of p53-Mdm2 dynamics has recently been provided by \cite{Bi:2015}. For a more complete understanding of how PDCD5 functions to regulate DNA repair and apoptosis following DNA damage, the PDCD5 dynamics \cite{Zhuge2011} and the p53 pathway response need to be incorporated \cite{Zhang2009,Zhang2012}. These results are certainly important for additional studies that seek to improve our understanding of how recombinant human PDCD5 can be used in cancer treatment.

\section*{SUPPORTING MATERIAL}

\ack{An online supplement to this article can be found by visiting BJ Online at http://www.biophysj.org.}\vspace*{-3pt}

\section*{AUTHOR CONTRIBUTIONS}
J.L. and Y.C. designed the research; C.Z. performed the research; C.Z. and X.S. analyzed data; C.Z. draft the manuscript; C.Z., X.S. and J.L. edited and revised the manuscript; C.Z., X.S., Y.C. and J.L. approved the final version of the manuscript.

\section*{ACKNOWLEDGMENTS}

\ack{This work was supported by the Fundamental Research Funds for the Central Universities (NO. BLX2014-29) and the National Natural Science Foundation of China (11272169, 91430101, 31370898)}\vspace*{6pt}

\bibliographystyle{plain}

\end{document}